\documentclass[12pt]{article}
\begin{document}

\title{{\bf Excluding Black Hole Firewalls\\ with\\ Extreme Cosmic Censorship}
\thanks{Alberta-Thy-4-13, arXiv:1306.0562 [hep-th]}}

\author{
Don N. Page
\thanks{Internet address:
profdonpage@gmail.com}
\\
Department of Physics\\
4-183 CCIS\\
University of Alberta\\
Edmonton, Alberta T6G 2E1\\
Canada
}

\date{2013 March 13; revised 2014 May 21}

\maketitle
\large
\begin{abstract}
\baselineskip 25 pt

The AMPS argument for black hole firewalls seems to arise not only from the assumption of local effective field theory outside the stretched horizon but also from an overcounting of internal black hole states that include states that are singular in the past.  Here I propose to exclude such singular states by {\it Extreme Cosmic Censorship} (the conjectured principle that the universe is entirely nonsingular, except for transient singularities inside black and/or white holes).  I argue that the remaining set of nonsingular realistic states do not have firewalls but yet preserve information in Hawking radiation from black holes that form from nonsingular initial states.

\end{abstract}

\normalsize

\baselineskip 19 pt

\newpage

Almheiri, Marolf, Polchinski, and Sully (AMPS) \cite{AMPS} have given a provocative argument that suggests that an ``infalling observer burns up at the horizon'' of a sufficiently old black hole, so that the horizon becomes what they called a ``firewall.''  A brief form of the argument is the following:  The assumptions of unitary evolution and of local effective field theory outside the stretched horizon suggest that at late times the Hawking radiation is maximally entangled with what is just outside the remaining black hole.  This further suggests that what is just outside cannot be significantly entangled with what is just inside.  But without this latter entanglement, an observer falling into the black hole should be burned up by high-energy radiation moving along the horizon.

The AMPS argument has elicited a large number of responses (including a few \cite{Braunstein:2009my, Giddings:2011ks, Giddings:2012bm} to forms of the argument realized even before the AMPS paper), 
some of which appear to support the firewall idea 
\cite{Bousso:2012as, Susskind:2012rm, Susskind:2012uw, Giveon:2012kp, Saravani:2012is, Biermann:2013wz, Avery:2013exa, Giveon:2013ica, Smerlak:2013cha, Marolf:2013dba, Chowdhury:2013mka, Almheiri:2013wka, Bousso:2013wia, Hewitt:2013gfa, Kim:2013caa, Bousso:2013uka, Berenstein:2013tya, Park:2014mba, Silverstein:2014yza, Berenstein:2014pma}, 
others seem rather agnostic or ambivalent 
\cite{Bena:2012zi, Hwang:2012nn, Culetu:2012fh, Majhi:2013tw, Kim:2013fv, Park:2013rm, Lee:2013vga, Hartman:2013qma, Iizuka:2013yla, Chowdhury:2013tza, Shenker:2013pqa, VanRaamsdonk:2013sza, Brustein:2013ena, Gomes:2013bbl, Elvang:2013nva, Bena:2013dka, Shenker:2013yza, Avery:2013bea, Devin:2014sma, Carr:2014mya, Ong:2014maa, Moffat:2014aqa, Harlow:2014yoa}, 
and yet others of which raise skepticism about it
\cite{Braunstein:2009my, Giddings:2011ks, Giddings:2012bm, Nomura:2011rb, Nomura:2012sw, Mathur:2012jk, Chowdhury:2012vd, Banks:2012nn, Ori:2012jx, Brustein:2012jn, Hossenfelder:2012mr, Nomura:2012cx, Avery:2012tf, Larjo:2012jt, Rama:2012fm, Page:2012zc, Papadodimas:2012aq, Nomura:2012ex, Giddings:2012gc, Neiman:2012fx, Jacobson:2013ewa, Harlow:2013tf, Susskind:2013tg, Page:2013dx, Hsu:2013cw, Giddings:2013kcj, Gambini:2013ooa, Brustein:2013xga, Nomura:2013nya, Brustein:2013qma, Banks:2013cha, Lowe:2013zxa, Verlinde:2013uja, Verlinde:2013vja, Maldacena:2013xja, Hotta:2013clt, Mathur:2013gua, Torrieri:2013lwa, Gary:2013oja, Hutchinson:2013kka, Iizuka:2013kma, Germani:2013sra, Mathur:2013bra, Giddings:2013vda, Nomura:2013gna, Lloyd:2013bza, Hsu:2013fra, Hui:2013jfa, Mathur:2013qda, Giddings:2013noa, Papadodimas:2013wnh, Papadodimas:2013jku, Nomura:2013lia, Abramowicz:2013dla, Verlinde:2013qya, Ilgin:2013iba, Braunstein:2013mba, Susskind:2013lpa, Susskind:2014gsa, Akhoury:2013bia, Susskind:2013aaa, Brustein:2013uoa, Hossenfelder:2014jha, Brustein:2014faa, Golovnev:2014jua, Banks:2014xja, Freivogel:2014dca, Hawking:2014tga, Giddings:2014nla, Moffat:2014eua, Lowe:2014vfa, Susskind:2014rva, Hollowood:2014hta, Sasaki:2014spa, Varela:2014qua}.  
A rebuttal of many of the counterarguments has recently been made by Almheiri, Marolf, Polchinski, Sully, and Stanford (AMPSS) \cite{AMPSS}.

The {\it electrifying} {\it AMPS} paper is {\it currently} a {\it powerful} and {\it inflammatory} argument, but it is {\it potentially} so {\it shocking} that within the physics {\it circuit} there has {\it surged} considerable {\it impedance} against accepting its conclusion, which is in {\it high tension} with past beliefs, so some of us feel {\it charged} to put {\it energy} into {\it conducting} research to {\it resist} it and to try to develop a theoretical {\it firewall} to block the {\it flames} and protect the equivalence principle from the {\it blaze} of this {\it burning} issue in a {\it hot} topic.

The AMPS (and AMPSS) argument explicitly uses the assumption of ``low energy effective field theory valid beyond some microscopic distance from the horizon.''  Effective field theory is local, whereas the constraint equations of gravity are nonlocal, so the assumption of effective field theory is almost certainly incorrect.  However, the puzzle is how to incorporate the nonlocality of quantum gravity in a way that does not invalidate the observed approximate validity of local effective field theory.

Here I argue that the AMPS and AMPSS argument also implicitly uses quantum states inside the black hole that would be singular if evolved backward.  Furthermore, a counting of these states would not be bounded by anything like the exponential of the Bekenstein-Hawking entropy of the black hole (one-quarter the area in Planck units, $A/4$).  To get the right state counting for a black hole, the huge number of such past-singular states must be excluded.  Here I shall suggest excluding them by what I call {\bf Extreme Cosmic Censorship}:

{\it The universe is entirely nonsingular (except for singularities deep inside black holes and/or white holes which do not persist to the infinite future or past, with these singularities coming near the surface only when the holes have masses near the Planck mass that normally happens only close to the ends and/or beginnings of their lifetimes).}

I further argue that within the remaining allowed {\it `realistic'} quantum states, there are no firewalls, and the black hole information is preserved in the Hawking radiation of its evaporation, so that there can still be nearly maximal entanglement between an old black hole and the radiation it has already emitted.\footnote{By an old black hole, I mean a black hole (including its nearby surroundings) after the von Neumann entropy of the Hawking radiation it has already emitted (and which has left the nearby surroundings) has started to go down \cite{Page:2013dx}.  In this paper I am excluding the possibility of high-entropy black hole remnants and instead assuming that the dimension of the Hilbert space of realistic nonsingular quantum states for a black hole and its nearby surroundings is given approximately by the exponential of the Bekenstein-Hawking entropy, $\exp{(A/4)}$ for a black hole of area $A$ in Planck units.  For an old black hole, these black hole states will be nearly maximally entangled with a suitable subset of the early Hawking radiation states.}

Let me propose the following classification of possible states in quantum gravity and/or quantum cosmology:

\begin{enumerate}

\item {\it Unconstrained kinematic states} are elements of a general state space with no requirement that they obey the gravitational and other gauge constraint equations.  For example, in Wheeler-DeWitt quantum cosmology, they could be arbitrary wavefunctionals of three-geometries and matter field configurations on them, whether or not these wavefunctionals obey the Hamiltonian and momentum constraint equations.  Because such wavefunctionals can be varied independently for different parts of the three-geometries, they can be considered to have the ordinary quantum field theory property of locality.

\item {\it Constrained physical states} are elements of the the unconstrained kinematic states that obey the constraint equations.  For example, they could be wavefunctionals of the three-geometry and matter field configurations that obey the momentum constraints and the Wheeler-DeWitt equation.  Because such constrained wavefunctionals have the behavior in the interior of a three-geometry constrained by the asymptotic behavior of the gravitational field, and/or by other asymptotic quantities, they do not exhibit the full property of locality in the way that nongravitational quantum field theory does on a fixed globally hyperbolic background spacetime, in which one may vary the quantum state in a region without changing the expectation value of any operator confined to any spacelike separated region.  (For example, if one has suitable asymptotically anti-de Sitter boundary conditions that give a unique constrained physical quantum state for each energy and angular momentum eigenvalue combination, then without changing the gravitational field at infinity, one cannot change the quantum state at all.)

\item {\it Nonsingular realistic states} are elements of the constrained physical states that obey {\it Extreme Cosmic Censorship} and do not have singularities other than transient black hole and/or white hole singularities.  In accord with what I am taking Extreme Cosmic Censorship to be, I am excluding singularities just inside the surface of a hole large in Planck units as a firewall would be.  Of course, when a hole has a size comparable to the Planck length, one would expect that the curvature at the surface would be Planckian, and in that case one might expect the singularity to be near the surface, perhaps being effectively naked for a small region of spacetime.  However, I mean for nonsingular realistic states to exclude states in which one can fall into a black hole from a region of curvature small in Planck units and suddenly see curvature large in Planck units or else a singularity, or the time reverse of this for white holes.  I am not counting the final evaporation of a black hole as a bad singularity to be excluded, since it is transient and might be confined to a point or small region of spacetime that perhaps quantum theory might heal.

\item {\it The actual state} is the actual quantum state of our entire universe or multiverse.  I am suggesting that it is one of the nonsingular realistic states.

\end{enumerate}

It is conceivable that all the constrained physical states do not have any singularities other than inside black holes that form and then evaporate away, in which case Extreme Cosmic Censorship would be enforced by the constraint equations.  However, in a spacetime picture it seems difficult to understand how this could be the case (though it could simply be a consequence of the inadequacy of a spacetime picture).  It would seem plausible that one should be able to perturb the inside of a black hole with a perturbation that at infinity has zero energy and angular momentum (since inside a black hole there are modes of both positive and negative energy), so that there is no change to the asymptotic gravitational field.  In fact, it na\"{\i}vely seems as if one should be able to make an infinite number of perturbations inside the black hole that do not affect the asymptotic behavior.  For example, at least classically there does not seem to be any obstruction to having any one of many different forms of firewalls just inside a black hole horizon but without affecting the outside at all.  Presumably one could also modify the full Kruskal geometry classically by having the horizons replaced by null singularities while leaving the geometry for $r > 2M$ exactly Schwarzschild.  I do not know how to count such states in quantum gravity but would suspect that there are infinitely many states that look like a black hole from the outside but allow arbitrary perturbations inside that classically are singular in either the past or future.  I think that it is only if one restricts to quantum states that come from smooth initial conditions without black holes (and evolve to smooth final states without black holes after the black holes evaporate) that one will get a finite set of black hole quantum states when the total energy is bounded.  One might then suppose that such restricted states that are smooth in the asymptotic past and future will not develop timelike or null singularities like firewalls.

Previous discussions of an apparently unbounded number of quantum states inside a boundary of fixed area include John Wheeler's `bag of gold' with a large cosmological region connected to an asymptotically flat spacetime through a small throat \cite{Wheeler}.  Rafael Sorkin, Robert Wald, and Zhen Jiu Zhang \cite{Sorkin:1981wd} have shown how this and other configurations can have arbitrarily large entropy within a fixed area, though they emphasize that these configurations will evolve to black holes in the future and white holes in the past and hence have both past and future singularities so that they ``cannot be built classically without starting from a white hole.''  More recently, Stephen Hsu and David Reeb have done further analyses of highly entropic objects, which they call `monsters' \cite{Hsu:2007dr,Hsu:2008yi,Hsu:2009kv}.  Yen Chin Ong and Pisin Chen show that some monsters are unstable but are not able to rule out all monsters \cite{Ong:2013mba}.

In this paper I shall make the hypothesis that not only is the set of constrained physical states a proper subset of the set of unconstrained kinematic states, but also the set of nonsingular realistic states is a proper subset of the set of constrained physical states, excluding `monster' states that have more entropy than black holes of the same area without violating the constraints of quantum gravity (and also `grireballs' that I myself have half-heartedly proposed \cite{Page:2012zc}).  In fact, I shall assume that the set of orthonormal constrained physical quantum states inside a black hole of bounded total mass and angular momentum at spatial infinity is infinite, whereas the analogous set of orthonormal nonsingular realistic quantum states is bounded and is approximately the exponential of the Bekenstein-Hawking entropy $S_\mathrm{BH} = A/4$, where $A$ is what the horizon area would be of a classical black hole with the same mass and angular momentum as that given by the gravitational field at infinity of the constrained physical state.  (I am ignoring phase-space factors for the location and motion of the black hole, which would be finite and much smaller than $A/4$ for a black hole large in Planck units in an asymptotically anti-de Sitter spacetime with a negative cosmological constant that is not exponentially close to zero.)

Another context in which the number of orthonormal constrained physical states appears to be infinitely larger than the number of orthonormal nonsingular realistic states is the cosmology of closed universes with a fixed positive cosmological constant.  It appears clear that if one takes a $k=+1$ Cauchy hypersurface at an arbitrarily late constant time in de Sitter spacetime, when the volume is arbitrarily large, one should be able to an arbitrarily large amount of entropy in arbitrarily many independent perturbations, at least in perturbation theory.  However, there have also been proposals to restrict the set of states to a finite number.

For example, Tom Banks \cite{Banks:2000fe} has proposed that the total number of quantum states needed to describe asymptotically de Sitter spacetime is the exponential of the Bekenstein-Hawking-Gibbons entropy, $\exp{(A/4)}$, where $A = 12\pi/\Lambda$ is the area of the cosmological event horizon of pure de Sitter spacetime.  He gave several arguments for this, including the argument that within the de Sitter horizon size the entropy apparently cannot be larger than $\exp{(3\pi/\Lambda)}$, and the argument that what is outside a single static patch are just gauge copies of what is inside one.  Banks recognizes that one can have initial conditions with larger entropy on a hypersurface of huge volume in the distant past of de Sitter spacetime, but he argues that most of these initial conditions will not lead to spacetimes that are asymptotically de Sitter in the distant future.  In other words, it appears that Banks is suggesting a restriction of the set of states with a positive cosmological constant to give a finite number.

(Incidentally, the Planck+highL+WMAP+BAO satellite and other data \cite{Ade:2013zuv} give the mean observed value of the asymptotic de Sitter entropy of our universe, $S_\mathrm{dS} = A/4 = 3\pi/\Lambda$, as being within 0.32\%, or within $1/11$ of a standard deviation of about 3.7\%, of $5^3\, 2^{400}$, which is a useful mnemonic, along with the fact that the present age of the universe is, within about $1/50$ of a standard deviation of about 2.7\%, $1.6\,\pi\, 2^{200}$ in Planck units, the fact that within $1/70$ of a standard deviation the cosmological constant $\Lambda$ is $1/(10\;{\mathrm{Gyr}})^2 = 10^{-20}\;$yr$^{-2}$, the fact that within $1/8$ of a standard deviation the cosmological constant is also {\it ten square attohertz} or $10\;\mathrm{aHz}^2 = 10^{-35}\;$s$^{-2}$, the fact that the present cosmic microwave background radiation (CMB) temperature is, within a bit less than one standard deviation of about 0.02\%, $2^{-100}/41$ in units of the Planck temperature $T_\mathrm{Planck} = (1.416\,833 \pm 0.000\,085)\times 10^{32}$ K, and the preliminary unchecked fact that the radiation entropy within our causal patch appears to be $S_\mathrm{rad} \approx 10 \times 2^{300} \approx (10/5^{9/4})\, S_\mathrm{dS}^{3/4}$ \cite{Scott:2013oib}.  Combining the first and third of these to set $5^3\, 2^{400} \approx 3\pi (10\;\mathrm{Gyr})^2$, with a year taken to be an average Julian year of 365.25 days, gives $t_\mathrm{Planck} \approx (12\pi/5)^{1/2}\, 2^{-200}\,\mathrm{Gyr} = 1.708\,76 \times 10^{-60}\,\mathrm{Gyr} = 5.392\,44 \times 10^{-44}$ seconds, which is about $1.000\,256 \pm 0.000\,059$ times the actual value for the Planck time, $t_\mathrm{Planck} = (5.391\,06 \pm 0.000\,32) \times 10^{-44}$ seconds, off by about 4.3 standard deviations, but probably good enough for most astrophysical applications.)

Raphael Bousso \cite{Bousso:2000nf} extends Banks' conjecture to give an `$N$ bound' on the number of degrees of freedom, $N = 3\pi/\Lambda$ (and an equal bound on the entropy), for all spacetimes with positive cosmological constant $\Lambda$.  He avoids the contradiction from an arbitrarily large number of perturbations on an arbitrarily large hypersurface of de Sitter spacetime by restricting the entropy to be the `observable entropy,' the entropy within the {\it causal diamond} that is the spacetime region with all points both to the future and to the past of an observer worldline.  The assumption is that only what is within one causal diamond is relevant for the corresponding observer, so that it is unnecessary for a quantum state giving predictions for that observer to include more.

If indeed each observation is sufficiently `local' that it can be given by what happens inside a causal diamond (say by having its measure or relative probability given by the expectation value of a quantum operator confined to the causal diamond \cite{Page:1995dc,Page:1995kw,Page:2000cs,Page:2001ba, Page:2008mx,Page:2008ns,Page:2009qe,Page:2011sr}), then one would only need the quantum state of the causal diamond (with different possibilities of what that causal diamond can be being different components of the quantum state of the causal diamond) to predict the measures or relative probabilities of all possible observations.  

Indeed, if all observations are made by observers with a bounded size that is much smaller than a causal diamond, as our brains are much smaller than the de Sitter radius, one would only need the quantum state in a region of that size, with all the different possibilities of what is inside that region being given by different components of the quantum state for the region.  Alternatively, though we do not know of examples, one might postulate that there could be observers larger than a causal diamond, so that the measure of their observations could be given by expectation values of operators that cannot be confined to a single causal diamond, such as the meta-observables considered by Edward Witten in his 2001 discussion of ``Quantum Gravity in de Sitter Space'' \cite{Witten:2001kn} that also considered the possibility that such quantum gravity might have a finite-dimensional Hilbert space.  As a result of these considerations, it is not clear to me that there is a preferred restriction to a causal diamond.

However, even if in principle one can deduce the measures of observations from a quantum state restricted to a causal diamond, one might find a simpler quantum state that includes many observer regions, or many causal diamonds.  Therefore, it is not obvious that for finding a simple quantum state of the universe it would be best to restrict attention to a single causal diamond, but it is certainly one approach that can be pursued.

In any case, Bousso is making an assumption that the relevant quantum states can be restricted to a finite number when there is a positive cosmological constant. Now there are counterexamples to the specific conjecture that he made in spacetime dimensions greater than four \cite{Bousso:2002fi} and also for Taub-Bolt(NUT) spacetimes in four dimensions \cite{Clarkson:2003kt,Clarkson:2004yp}.  However, it is still plausible that if one makes a suitable restriction of the quantum states, there may be only a finite number for a fixed positive value of the cosmological constant.  For example, Steven Giddings and Donald Marolf \cite{Giddings:2007nu} argue that when one does group averaging to get quantum states that fulfill the requirement that each be de Sitter invariant, then when one restricts to states that are asymptotically de Sitter in both the future and the past, the set of such states is finite dimensional.  They further conjecture that the number of nonperturbative asymptotically de Sitter states is the exponential of the Bekenstein-Hawking-Gibbons entropy $\exp{(3\pi/\Lambda)}$.

Besides the arbitrarily large number of states for de Sitter spacetime perturbed at arbitrarily late or early times when the spatial volume is arbitrarily large and when one does not require that the spacetime be asymptotically de Sitter in both the future and the past (see, for example, \cite{Witten:2001kn} for a nice discussion of this point as well as pointing out that it need not be true for nonperturbative quantum gravity), one could alternatively add infinitely many perturbations to the unwrapped Nariai metric (the covering space of $S^2\times dS_2$, which has an infinitely long time-symmetric throat) \cite{Page:2007hp, Page:2009ct}.

Nevertheless, if one evolves backward in time de Sitter with arbitrarily many perturbations added at arbitrarily late times, or perturbations to the infinitely long unwrapped Nariai metric, the Einstein equations would probably lead to a big-bang and/or big-crunch singularity.  Even in these cases, there can be asymptotically locally de Sitter regions to the past and future, so for a finite number of states it is not sufficient to require that there exist one or more asymptotically de Sitter regions in both the past and future.  For getting a finite number of states, in \cite{Page:2007hp} I proposed excluding states that have a big bang or big crunch or which split into multiple asymptotic de Sitter spacetimes as the Nariai metric would with a large class of perturbations.  I conjecture that this single-nonsingular-de Sitter restriction would lead to a finite number of quantum states \cite{Page:2007hp, Page:2009ct}.

In \cite{Page:2011yd}, I showed that the total canonical (Liouville-Henneaux-Gibbons-Hawking-Stewart) measure is finite for completely nonsingular Friedmann-Lema\^{\i}tre-Robertson-Walker classical universes with a minimally coupled massive scalar field and a positive cosmological constant.  This suggests that the number of nonsingular quantum states may also be finite.

Similar considerations suggest that black hole states that are nonsingular to the asymptotic past and future (before and after all transient black hole and/or white hole singularities have formed and disappeared) may give a finite number of orthonormal quantum nonsingular realistic states when the mass and effective volume for the location are bounded (say with an upper bound on the energy and a small negative cosmological constant to give asymptotically anti-de Sitter spacetime with an effectively finite volume, thus avoiding an infinite phase space for the location of the black hole).  One would expect this finite number to be given approximately by the exponential of the Bekenstein-Hawking entropy, one-quarter the area of the largest static black hole with mass equal to the upper bound on the energy with the fixed cosmological constant.

The firewalls proposed in the AMPS and AMPSS papers \cite{AMPS, AMPSS} appear to be particular examples of violations of Extreme Cosmic Censorship.  I am proposing that firewall states are just singular states that one should exclude from being considered physically realistic, in a way similar to the way Horowitz and Myers \cite{Horowitz:1995ta} suggested that the naked singularity of negative-mass Schwarzschild has the value of providing us with a criterion for eliminating such solutions and for giving a stable ground state.  Even if one considers a firewall of large but finite energy density, evolving it backward would seem to lead to a singularity violating Extreme Cosmic Censorship.

Long after the previous paragraph was originally written, Stephen Hawking \cite{Hawking-2014-04-07} told James Hartle, Thomas Hertog, and me the following:  ``If black holes leave remnants, excited-AdS must also start with remnants.  So let's assume black holes don't leave remnants.  Then holes will grow to a maximum size absorbing information, and shrink to zero emitting radiation.  Firewalls are like remnants.  If you have firewalls, excited-AdS will be initially singular.''

Using Extreme Cosmic Censorship to restrict to nonsingular realistic states would exclude unconstrained kinematic states and constrained physical states that within those broader spaces of states do not have strong entanglement between what is just outside and what is just inside the black hole horizon.  Thus it na\"{\i}vely appears to require near-maximal entanglement between subsystem $A$ that is the near-horizon interior of the black hole and subsystem $B$ that is the near-horizon exterior of the hole.  Under the assumption of effective local field theory outside the black hole so that $B$ evolves unitarily to become the late Hawking radiation, the principle of quantum monogamy \cite{monogamy} implies that the near-maximal entanglement of $B$ with $A$ would prevent $B$ from being also nearly maximally entangled with the early Hawking radiation $R$ as would be needed for the black hole evaporation to lead to a nearly pure final state and preserve unitarity for the formation of the black hole in a nearly pure initial state \cite{AMPS}.  This is the essence of the argument of AMPS that unitarity plus locality outside the black hole requires a breaking of the strong entanglement between $A$ just inside and $B$ just outside the black hole and hence the formation of a firewall as seen by an infalling observer.

However, within the restrictions of Extreme Cosmic Censorship to nonsingular realistic states, there is no freedom for $A$ and $B$ to appear to have unentangled degrees of freedom that would give a firewall.  Within the restricted class of states, there are no such unentangled states.  With no such unentangled nonsingular realistic states, the concept of entanglement between $A$ and $B$ becomes rather illusory, an artifact of shifting attention to a larger class of singular unrealistic states.  Within the set of nonsingular realistic states, the states across the horizon are whatever smooth states they are, without their actually exhibiting entanglement between truly independent nonsingular realistic degrees of freedom.

If the assumption of local effective field theory outside the stretched horizon were valid, one could give the rebuttal that $B$ evolves unitarily to become the late Hawking radiation, which must then be highly entangled with the early Hawking radiation $R$ for the entire process of black-hole evaporation to be unitary.  This would then prevent $B$ from having even the illusory entanglement with $A$.  My answer would be that I am not claiming that firewalls can be avoided while retaining the other three AMPS postulates, but that indeed one should give up the assumption of local effective field theory.  

Indeed there presumably must be some process that
transfers the entanglement between $R$ (the early Hawking radiation) and
the ($A$,$B$) system (the remaining black hole and its nearby
surroundings) to an entanglement between $B$ (once these modes leave the
region near the black hole to become part of the late Hawking
radiation) and the early Hawking radiation $R$, and this would almost certainly involve violations of local effective field theory, perhaps of the form considered by Steven Giddings \cite{Giddings:2012gc, Giddings:2013kcj, Giddings:2013vda, Giddings:2013noa, Giddings:2014nla}.  The main point of the present paper is that within the context of giving up the assumption of local effective field theory, the proposal of Extreme Cosmic Censorship may provide a simple way for excluding firewalls.

As an analogy for the distinction between constrained physical states and nonsingular realistic states, consider the singlet state of two spin-half particles.  Normally, restricting to the analogue of nonsingular realistic states, we say this state has maximal entanglement between the one spin and the other.  However, if we considered a small mathematical sphere around each particle (small compared with the particle separation but very large in Planck units) and considered the space of states in which the fields just inside and just outside each sphere are kinematically independent (analogous to the set of constrained physical states), there would be huge entanglements (though finite if one has a finite cutoff, say at the Planck length) across each of the spheres, so that what is inside one sphere containing one of the two spin-half particles would not be considered at all nearly maximally entangled with what is inside the other sphere containing the other spin-half particle.  So in this way of considering the degrees of freedom, the two spin-half particles and their respective spheres would have only a relatively small amount of entanglement.  (The spins would be entangled, but not most of the other degrees of freedom within the respective spheres.)  The joint system of the insides of those two spheres that are much further apart than their radii (which are themselves much greater than the cutoff length used to regulate the entanglement between the insides and outsides of each of the spheres) would be far from being in a pure quantum state.

However, in the more usual way of looking at the two spin-half particles, we exclude the high-energy states that would correspond to allowing the quantum fields on the opposite sides of the spheres to be unentangled.  Therefore, we do not count this entanglement in the larger set of states that include the very high-energy states (energies up to at least Planck energies with a Planck-scale length cutoff).  We can say that with respect to the low-energy states (energies not much larger than the sum of the rest masses of the two spin-half particles), the entanglement between the insides and the outsides of the respective spheres is illusory, so that indeed we can consider the two spins to be maximally (or very nearly maximally) entangled.

In a similar way, when we restrict to nonsingular realistic states in quantum gravity that obey Extreme Cosmic Censorship, we do not count the huge na\"{\i}ve entanglement between field degrees of freedom on opposite sides of black hole horizons.  This entanglement in a larger space of quantum states is illusory with respect to the nonsingular realistic states, so that within the nonsingular realistic states, an old black hole and its nearby environs can be very nearly maximally entangled with the distant Hawking radiation that has been emitted earlier, so that the total black hole formation and evaporation process can be unitary without violating quantum monogamy.

As the revisions of this paper were being finished, Leonard Susskind sent me an email saying, ``I think we are on the same Page,'' and noting that Extreme Cosmic Censorship is closely related to the `stretching criterion' in his recent papers on ``Computational Complexity and Black Hole Horizons'' \cite{Susskind:2014rva,Susskind:2014ira}:  ``{\it Black holes are formed in such a way that the complexity of the state increases.}''  The main difference is that Susskind allows complexity to decrease in rare circumstances, analogous to violations of the second law of thermodynamics, whereas my Extreme Cosmic Censorship is meant to be put forward as an absolute restriction in the set of realistic quantum states.  However, since I do not have a precise criterion of what nonsingular states are in quantum gravity, it might well be that one cannot make an absolute restriction that I would have hoped Extreme Cosmic Censorship could be.

In conclusion, when one uses Extreme Cosmic Censorship (which excludes big bang and big crunch singularities as well as naked singularities, that is, all singularities other than transient black hole singularities) to restrict the allowed set of states to nonsingular realistic quantum states, the apparent strong entanglement across black hole event horizons is an illusion from a different viewpoint of a much larger space of states.  There is no conflict between this illusory `entanglement' and black hole formation and evaporation without loss of information, though this argument does not imply that one can avoid giving up the assumption of local effective field theory outside some microscopic distance from the black hole horizon, an assumption that has long appeared to be dubious.

I have benefited from Leonard Susskind's hospitality at the 2012 November 30 -- December 1 firewall conference at Stanford University, from discussions there and by email, and from other communications with and from the participants and others, including, but not limited to, the following names who come to mind at present:  Andy Albrecht, Ahmed Almheiri, Steve Avery, Tom Banks, Sam Braunstein, Raphael Bousso, Adam Brown, Willy Fischler, Ben Freivogel, Gary Gibbons, Steve Giddings, Daniel Harlow, Stephen Hawking, Patrick Hayden, Simeon Hellerman, Gary Horowitz, Viqar Husain, Ted Jacobson, Shamit Kachru, Matt Kleban, Kayll Lake, Stefan Leichenauer, Juan Maldacena, Don Marolf, Samir Mathur, Yasunori Nomura, Joe Polchinski, John Preskill, Mark van Raamsdonk, Steve Shenker, Eva Silverstein, Mark Srednicki, Rafael Sorkin, Douglas Stanford, Dejan Stojkovic, Andy Strominger, Jamie Sully, Lenny Susskind, David Turton, Bill Unruh, Erik Verlinde, Herman Verlinde, Aron Wall, Nick Warner, Ed Witten, and Karol \.{Z}yczkowski.  After a previous version of this paper was posted on the arXiv, I am grateful for discussions with Raphael Bousso, Steven Carlip, Gary Gibbons, Steve Giddings, Daniel Harlow, Jim Hartle, Stephen Hawking, Thomas Hertog, Don Marolf, Joe Polchinski, Lenny Susskind, and an anonymous JCAP referee (most probably one of those above I have already acknowledged, but he or she deserves an extra acknowledgment for emphasizing that I should make it clear that my proposal does not avoid the need for nonlocality) which helped me with adding references and making other revisions of the paper.  The discussions with Stephen Hawking and a few others were enabled by the gracious hospitality of the Mitchell family and Texas A \& M University at a workshop at Great Brampton House, Herefordshire, England.  This work was supported in part by the Natural Sciences and Engineering Research Council of Canada.


\newpage

\baselineskip 4pt

\begin{thebibliography}{99}

\bibitem{AMPS} 
  A.~Almheiri, D.~Marolf, J.~Polchinski and J.~Sully,
  ``Black Holes: Complementarity or Firewalls?,''
  JHEP {\bf 1302}, 062 (2013)
  [arXiv:1207.3123 [hep-th]];
  cf.\ S.~L.~Braunstein, S.~Pirandola and K.~\.{Z}yczkowski,
  ``Entangled Black Holes as Ciphers of Hidden Information,''
  Physical Review Letters {\bf 110}, 101301S (2013)
  [arXiv:0907.1190 [quant-ph]]
  for a similar prediction from different assumptions.

\bibitem{Braunstein:2009my} 
  S.~L.~Braunstein, S.~Pirandola and K.~\.{Z}yczkowski,
  ``Better Late than Never: Information Retrieval from Black Holes,''
  Phys.\ Rev.\ Lett.\  {\bf 110}, no. 10, 101301 (2013)
  [arXiv:0907.1190 [quant-ph]].
    
\bibitem{Giddings:2011ks} 
  S.~B.~Giddings,
  ``Models for Unitary Black Hole Disintegration,''
  Phys.\ Rev.\ D {\bf 85}, 044038 (2012)
  [arXiv:1108.2015 [hep-th]].
  
\bibitem{Giddings:2012bm} 
  S.~B.~Giddings,
  ``Black Holes, Quantum Information, and Unitary Evolution,''
  Phys.\ Rev.\ D {\bf 85}, 124063 (2012)
  [arXiv:1201.1037 [hep-th]].

\bibitem{Bousso:2012as} 
  R.~Bousso,
  ``Complementarity Is Not Enough,''
  Phys.\ Rev.\ D {\bf 87}, no. 12, 124023 (2013)
  [arXiv:1207.5192 [hep-th]].
    
\bibitem{Susskind:2012rm} 
  L.~Susskind,
  ``Singularities, Firewalls, and Complementarity,''
  arXiv:1208.3445 [hep-th]. 
  
\bibitem{Susskind:2012uw} 
  L.~Susskind,
  ``The Transfer of Entanglement: The Case for Firewalls,''
  arXiv:1210.2098 [hep-th]. 

\bibitem{Giveon:2012kp} 
  A.~Giveon and N.~Itzhaki,
  ``String Theory Versus Black Hole Complementarity,''
  JHEP {\bf 1212}, 094 (2012)
  [arXiv:1208.3930 [hep-th]].  

\bibitem{Saravani:2012is} 
  M.~Saravani, N.~Afshordi and R.~B.~Mann,
  ``Empty Black Holes, Firewalls, and the Origin of Bekenstein-Hawking Entropy,''
  arXiv:1212.4176 [hep-th].

\bibitem{Biermann:2013wz} 
  P.~L.~Biermann and B.~C.~Harms,
  ``A Comprehensive Model of Dark Energy, Inflation and Black Holes,''
  arXiv:1302.0040 [gr-qc].

\bibitem{Avery:2013exa} 
  S.~G.~Avery and B.~D.~Chowdhury,
  ``Firewalls in AdS/CFT,''
  arXiv:1302.5428 [hep-th].

\bibitem{Giveon:2013ica} 
  A.~Giveon and N.~Itzhaki,
  ``String Theory at the Tip of the Cigar,''
  JHEP {\bf 1309}, 079 (2013)
  [arXiv:1305.4799 [hep-th]].

\bibitem{Smerlak:2013cha} 
  M.~Smerlak,
  ``The Two Faces of Hawking Radiation,''
  Int.\ J.\ Mod.\ Phys.\ D {\bf 22}, 1342019 (2013)
  [arXiv:1307.2227 [gr-qc]].

\bibitem{Marolf:2013dba} 
  D.~Marolf and J.~Polchinski,
  ``Gauge/Gravity Duality and the Black Hole Interior,''
  Phys.\ Rev.\ Lett.\  {\bf 111}, 171301 (2013)
  [arXiv:1307.4706 [hep-th]].
  
\bibitem{Chowdhury:2013mka} 
  B.~D.~Chowdhury,
  ``Cool Horizons Lead to Information Loss,''
  JHEP {\bf 1310}, 034 (2013)
  [arXiv:1307.5915 [hep-th]].
  
\bibitem{Almheiri:2013wka} 
  A.~Almheiri and J.~Sully,
  ``An Uneventful Horizon in Two Dimensions,''
  JHEP {\bf 1402}, 108 (2014)
  [arXiv:1307.8149 [hep-th]].  

\bibitem{Bousso:2013wia} 
  R.~Bousso,
  ``Firewalls from Double Purity,''
  Phys.\ Rev.\ D {\bf 88}, 084035 (2013)
  [arXiv:1308.2665 [hep-th]].  

\bibitem{Hewitt:2013gfa} 
  M.~Hewitt,
  ``Thermal Duality and Gravitational Collapse in Heterotic String Theories,''
  arXiv:1309.7578 [hep-th].  

\bibitem{Kim:2013caa} 
  W.~Kim and E.~J.~Son,
  ``Freely Falling Observer and Black Hole Radiation,''
  Mod.\ Phys.\ Lett.\ A {\bf 29}, 1450052 (2014)
  [arXiv:1310.1458 [hep-th]].

\bibitem{Bousso:2013uka} 
  R.~Bousso and D.~Stanford,
  ``Measurements without Probabilities in the Final State Proposal,''
  Phys.\ Rev.\ D {\bf 89}, 044038 (2014)
  [arXiv:1310.7457 [hep-th]].

\bibitem{Berenstein:2013tya} 
  D.~Berenstein and E.~Dzienkowski,
  ``Numerical Evidence for Firewalls,''
  arXiv:1311.1168 [hep-th].

\bibitem{Park:2014mba} 
  I.~Y.~Park,
  ``Indication for Unsmooth Horizon Induced by Quantum Gravity Interaction,''
  arXiv:1401.1492 [hep-th].

\bibitem{Silverstein:2014yza} 
  E.~Silverstein,
  ``Backdraft: String Creation in an Old Schwarzschild Black Hole,''
  arXiv:1402.1486 [hep-th].

\bibitem{Berenstein:2014pma} 
  D.~Berenstein,
  ``Sketches of Emergent Geometry in the Gauge/Gravity Duality,''
  arXiv:1404.7052 [hep-th].

\bibitem{Bena:2012zi} 
  I.~Bena, A.~Puhm and B.~Vercnocke,
  ``Non-Extremal Black Hole Microstates: Fuzzballs of Fire or Fuzzballs of Fuzz ?,''
  JHEP {\bf 1212}, 014 (2012)
  [arXiv:1208.3468 [hep-th]].

\bibitem{Hwang:2012nn} 
  D.-i.~Hwang, B.-H.~Lee and D.-h.~Yeom,
  ``Is the Firewall Consistent?: Gedanken Experiments on Black Hole Complementarity and Firewall Proposal,''
  JCAP {\bf 1301}, 005 (2013)
  [arXiv:1210.6733 [gr-qc]].

\bibitem{Culetu:2012fh} 
  H.~Culetu,
  ``Comment on 'Empty Black Holes, Firewalls, and the Origin of Bekenstein-Hawking Entropy',''
  arXiv:1212.6877 [hep-th].

\bibitem{Majhi:2013tw} 
  A.~Majhi and P.~Majumdar,
  ``Quantum Hairs and Isolated Horizon Entropy from Chern-Simons Theory,''
  arXiv:1301.4553 [gr-qc].

\bibitem{Kim:2013fv} 
  W.~Kim, B.-H.~Lee and D.-h.~Yeom,
  ``Black hole Complementarity and Firewall in Two Dimensions,''
  JHEP {\bf 1305}, 060 (2013)
  [arXiv:1301.5138 [gr-qc]].

\bibitem{Park:2013rm} 
  I.~Y.~Park,
  ``On the Pattern of Black Hole Information Release,''
  Int.\ J.\ Mod.\ Phys.\ A {\bf 29}, 1450047 (2014)
  [arXiv:1301.6320 [hep-th]].

\bibitem{Lee:2013vga} 
  B.-H.~Lee and D.-h.~Yeom,
  ``Status Report: Black Hole Complementarity Controversy,''
  Nucl.\ Phys.\ Proc.\ Suppl.\  {\bf 246-247}, 178 (2014)
  [arXiv:1302.6006 [gr-qc]].

\bibitem{Hartman:2013qma} 
  T.~Hartman and J.~Maldacena,
  ``Time Evolution of Entanglement Entropy from Black Hole Interiors,''
  JHEP {\bf 1305}, 014 (2013)
  [arXiv:1303.1080 [hep-th]].

\bibitem{Iizuka:2013yla} 
  N.~Iizuka, D.~Kabat, S.~Roy and D.~Sarkar,
  ``Black Hole Formation at the Correspondence Point,''
  Phys.\ Rev.\ D {\bf 87}, no. 12, 126010 (2013)
  [arXiv:1303.7278 [hep-th]].

\bibitem{Chowdhury:2013tza} 
  B.~D.~Chowdhury,
  ``Black Holes versus Firewalls and Thermo-Field Dynamics,''
  Int.\ J.\ Mod.\ Phys.\ D {\bf 22}, 1342011 (2013)
  [arXiv:1305.6343 [gr-qc]].

\bibitem{Shenker:2013pqa} 
  S.~H.~Shenker and D.~Stanford,
  ``Black Holes and the Butterfly Effect,''
  JHEP {\bf 1403}, 067 (2014)
  [arXiv:1306.0622 [hep-th]].

\bibitem{VanRaamsdonk:2013sza} 
  M.~Van Raamsdonk,
  ``Evaporating Firewalls,''
  arXiv:1307.1796 [hep-th].

\bibitem{Brustein:2013ena} 
  R.~Brustein and A.~J.~M.~Medved,
  ``Phases of Information Release During Black Hole Evaporation,''
  JHEP {\bf 1402}, 116 (2014)
  [arXiv:1310.5861 [hep-th], arXiv:1310.5861].

\bibitem{Gomes:2013bbl} 
  H.~Gomes and G.~Herczeg,
  ``A Rotating Black Hole Solution for Shape Dynamics,''
  arXiv:1310.6095 [gr-qc].

\bibitem{Elvang:2013nva} 
  H.~Elvang and G.~T.~Horowitz,
  ``Quantum Gravity via Supersymmetry and Holography,''
  arXiv:1311.2489 [gr-qc].  

\bibitem{Bena:2013dka} 
  I.~Bena and N.~P.~Warner,
  ``Resolving the Structure of Black Holes: Philosophizing with a Hammer,''
  arXiv:1311.4538 [hep-th].

\bibitem{Brustein:2013uoa} 
  R.~Brustein and A.~J.~M.~Medved,
  ``Horizons of Semiclassical Black Holes are Cold,''
  arXiv:1312.0880 [hep-th].  

\bibitem{Shenker:2013yza} 
  S.~H.~Shenker and D.~Stanford,
  ``Multiple Shocks,''
  arXiv:1312.3296 [hep-th].  

\bibitem{Avery:2013bea} 
  S.~G.~Avery and B.~D.~Chowdhury,
  ``No Holography for Eternal AdS Black Holes,''
  arXiv:1312.3346 [hep-th].  

\bibitem{Devin:2014sma} 
  M.~Devin,
  ``Musings on Firewalls and the Information Paradox,''
  arXiv:1401.0588 [gr-qc].

\bibitem{Carr:2014mya} 
  B.~J.~Carr,
  ``The Black Hole Uncertainty Principle Correspondence,''
  arXiv:1402.1427 [gr-qc].

\bibitem{Ong:2014maa} 
  Y.~C.~Ong, B.~McInnes and P.~Chen,
  ``Why Hawking Radiation Cannot Be Decoded,''
  arXiv:1403.4886 [hep-th].

\bibitem{Moffat:2014aqa} 
  J.~W.~Moffat and V.~T.~Toth,
  ``Karlhede's Invariant and the Black Hole Firewall Proposal,''
  arXiv:1404.1845 [gr-qc].

\bibitem{Harlow:2014yoa} 
  D.~Harlow,
  ``Aspects of the Papadodimas-Raju Proposal for the Black Hole Interior,''
  arXiv:1405.1995 [hep-th].

\bibitem{Nomura:2011rb} 
  Y.~Nomura,
  ``Quantum Mechanics, Spacetime Locality, and Gravity,''
  Found.\ Phys.\  {\bf 43}, 978 (2013)
  [arXiv:1110.4630 [hep-th]].
  
\bibitem{Nomura:2012sw} 
  Y.~Nomura, J.~Varela and S.~J.~Weinberg,
  ``Complementarity Endures: No Firewall for an Infalling Observer,''
  JHEP {\bf 1303}, 059 (2013) [arXiv:1207.6626 [hep-th]].  

 \bibitem{Mathur:2012jk} 
  S.~D.~Mathur and D.~Turton,
  ``Comments on Black Holes I: The Possibility of Complementarity,''
  JHEP {\bf 1401}, 034 (2014)
  [arXiv:1208.2005 [hep-th]]. 

\bibitem{Chowdhury:2012vd} 
  B.~D.~Chowdhury and A.~Puhm,
  ``Is Alice burning or fuzzing?,''
  Phys.\ Rev.\ D {\bf 88}, 063509 (2013)
  [arXiv:1208.2026 [hep-th]].  

\bibitem{Banks:2012nn} 
  T.~Banks and W.~Fischler,
  ``Holographic Space-Time Does Not Predict Firewalls,''
  arXiv:1208.4757 [hep-th].  

\bibitem{Ori:2012jx} 
  A.~Ori,
  ``Firewall or Smooth Horizon?,''
  arXiv:1208.6480 [gr-qc]. 

\bibitem{Brustein:2012jn} 
  R.~Brustein,
  ``Origin of the Blackhole Information Paradox,''
  Fortsch.\ Phys.\  {\bf 62}, 255 (2014)
  [arXiv:1209.2686 [hep-th]].

\bibitem{Hossenfelder:2012mr} 
  S.~Hossenfelder,
  ``Comment on the Black Hole Firewall,''
  arXiv:1210.5317 [gr-qc].

\bibitem{Nomura:2012cx} 
  Y.~Nomura, J.~Varela and S.~J.~Weinberg,
  ``Black Holes, Information, and Hilbert Space for Quantum Gravity,''
  Phys.\ Rev.\ D {\bf 87}, no. 8, 084050 (2013)
  [arXiv:1210.6348 [hep-th]].

\bibitem{Avery:2012tf} 
  S.~G.~Avery, B.~D.~Chowdhury and A.~Puhm,
  ``Unitarity and Fuzzball Complementarity: 'Alice Fuzzes but May not Even Know It!',''
  JHEP {\bf 1309}, 012 (2013)
  [arXiv:1210.6996 [hep-th]].

\bibitem{Larjo:2012jt} 
  K.~Larjo, D.~A.~Lowe and L.~Thorlacius,
  ``Black Holes without Firewalls,''
  Phys.\ Rev.\ D {\bf 87}, no. 10, 104018 (2013)
  [arXiv:1211.4620 [hep-th]].

\bibitem{Rama:2012fm} 
  S.~K.~Rama,
  ``Remarks on Black Hole Evolution a la Firewalls and Fuzzballs,''
  arXiv:1211.5645 [hep-th].

\bibitem{Page:2012zc} 
  D.~N.~Page,
  ``Hyper-Entropic Gravitational Fireballs (Grireballs) with Firewalls,''
  JCAP {\bf 1304}, 037 (2013)
  [arXiv:1211.6734 [hep-th]].

\bibitem{Papadodimas:2012aq} 
  K.~Papadodimas and S.~Raju,
  ``An Infalling Observer in AdS/CFT,''
  JHEP {\bf 1310}, 212 (2013)
  [arXiv:1211.6767 [hep-th]].

\bibitem{Nomura:2012ex} 
  Y.~Nomura and J.~Varela,
  ``A Note on (No) Firewalls: The Entropy Argument,''
  JHEP {\bf 1307}, 124 (2013)
  [arXiv:1211.7033 [hep-th]].

\bibitem{Giddings:2012gc} 
  S.~B.~Giddings,
  ``Nonviolent Nonlocality,''
  Phys.\ Rev.\ D {\bf 88}, 064023 (2013)
  [arXiv:1211.7070 [hep-th]].

\bibitem{Neiman:2012fx} 
  Y.~Neiman,
  ``On-Shell Actions with Lightlike Boundary Data,''
  arXiv:1212.2922 [hep-th].

\bibitem{Jacobson:2013ewa} 
  T.~Jacobson,
  ``Boundary Unitarity and the Black Hole Information Paradox,''
  Int.\ J.\ Mod.\ Phys.\ D {\bf 22}, 1342002 (2013)
  [arXiv:1212.6944 [hep-th]].  

\bibitem{Harlow:2013tf} 
  D.~Harlow and P.~Hayden,
  ``Quantum Computation vs. Firewalls,''
  JHEP {\bf 1306}, 085 (2013)
  [arXiv:1301.4504 [hep-th]].

\bibitem{Susskind:2013tg} 
  L.~Susskind,
  ``Black Hole Complementarity and the Harlow-Hayden Conjecture,''
  arXiv:1301.4505 [hep-th].

\bibitem{Page:2013dx} 
  D.~N.~Page,
  ``Time Dependence of Hawking Radiation Entropy,''
  JCAP {\bf 1309}, 028 (2013)
  [arXiv:1301.4995 [hep-th]].

\bibitem{Hsu:2013cw} 
  S.~D.~H.~Hsu,
  ``Macroscopic Superpositions and Black Hole Unitarity,''
  arXiv:1302.0451 [hep-th].

\bibitem{Giddings:2013kcj} 
  S.~B.~Giddings,
  ``Nonviolent Information Transfer from Black Holes: A Field Theory Parametrization,''
  Phys.\ Rev.\ D {\bf 88}, no. 2, 024018 (2013)
  [arXiv:1302.2613 [hep-th]].

\bibitem{Gambini:2013ooa} 
  R.~Gambini and J.~Pullin,
  ``Loop Quantization of the Schwarzschild Black Hole,''
  Phys.\ Rev.\ Lett.\  {\bf 110}, no. 21, 211301 (2013)
  [arXiv:1302.5265 [gr-qc]].

\bibitem{Brustein:2013xga} 
  R.~Brustein and A.~J.~M.~Medved,
  ``Semiclassical Black Holes Expose Forbidden Charges and Censor Divergent Densities,''
  JHEP {\bf 1309}, 108 (2013)
  [arXiv:1302.6086 [hep-th]].

\bibitem{Nomura:2013nya} 
  Y.~Nomura, J.~Varela and S.~J.~Weinberg,
  ``Low Energy Description of Quantum Gravity and Complementarity,''
  arXiv:1304.0448 [hep-th].

\bibitem{Brustein:2013qma} 
  R.~Brustein and A.~J.~M.~Medved,
  ``Restoring Predictability in Semiclassical Gravitational Collapse,''
  JHEP {\bf 1309}, 015 (2013)
  [arXiv:1305.3139 [hep-th]].

\bibitem{Banks:2013cha} 
  T.~Banks and W.~Fischler,
  ``No Firewalls in Holographic Space-Time or Matrix Theory,''
  arXiv:1305.3923 [hep-th].

\bibitem{Lowe:2013zxa} 
  D.~A.~Lowe and L.~Thorlacius,
  ``Pure States and Black Hole Complementarity,''
  Phys.\ Rev.\ D {\bf 88}, 044012 (2013)
  [arXiv:1305.7459 [hep-th]].

\bibitem{Verlinde:2013uja} 
  E.~Verlinde and H.~Verlinde,
  ``Passing through the Firewall,''
  arXiv:1306.0515 [hep-th].

\bibitem{Verlinde:2013vja} 
  E.~Verlinde and H.~Verlinde,
  ``Black Hole Information as Topological Qubits,''
  arXiv:1306.0516 [hep-th].

\bibitem{Maldacena:2013xja} 
  J.~Maldacena and L.~Susskind,
  ``Cool Horizons for Entangled Black Holes,''
  Fortsch.\ Phys.\  {\bf 61}, 781 (2013)
  [arXiv:1306.0533 [hep-th]].

\bibitem{Hotta:2013clt} 
  M.~Hotta, J.~Matsumoto and K.~Funo,
  ``Black Hole Firewalls Require Huge Energy of Measurement,''
  arXiv:1306.5057 [quant-ph].

\bibitem{Mathur:2013gua} 
  S.~D.~Mathur and D.~Turton,
  ``The Flaw in the Firewall Argument,''
  arXiv:1306.5488 [hep-th].
  
\bibitem{Torrieri:2013lwa} 
  G.~Torrieri,
  ``Multi-Particle Correlations, Many Particle Systems, and Entropy in Effective Field Theories,''
  arXiv:1306.5719 [hep-th].  

\bibitem{Gary:2013oja} 
  M.~Gary,
  ``Still No Rindler Firewalls,''
  arXiv:1307.4972 [hep-th].

\bibitem{Hutchinson:2013kka} 
  J.~Hutchinson and D.~Stojkovic,
  ``Icezones Instead of Firewalls: Extended Entanglement Beyond the Event Horizon and Unitary Evaporation of a Black Hole,''
  arXiv:1307.5861 [hep-th].

\bibitem{Iizuka:2013kma} 
  N.~Iizuka and S.~Terashima,
  ``Brick Walls for Black Holes in AdS/CFT,''
  arXiv:1307.5933 [hep-th].

\bibitem{Germani:2013sra} 
  C.~Germani,
  ``On the Many Saddle Points Description of Quantum Black Holes,''
  arXiv:1307.6238 [hep-th].

\bibitem{Mathur:2013bra} 
  S.~D.~Mathur,
  ``What Happens at the Horizon?,''
  Int.\ J.\ Mod.\ Phys.\ D {\bf 22}, 1341016 (2013)
  [arXiv:1308.2785 [hep-th]].

\bibitem{Giddings:2013vda} 
  S.~B.~Giddings,
  ``Statistical Physics of Black Holes as Quantum-Mechanical Systems,''
  Phys.\ Rev.\ D {\bf 88}, 104013 (2013)
  [arXiv:1308.3488 [hep-th]].

\bibitem{Nomura:2013gna} 
  Y.~Nomura, J.~Varela and S.~J.~Weinberg,
  ``Black Holes or Firewalls: A Theory of Horizons,''
  Phys.\ Rev.\ D {\bf 88}, 084052 (2013)
  [arXiv:1308.4121 [hep-th]].

\bibitem{Lloyd:2013bza} 
  S.~Lloyd and J.~Preskill,
  ``Unitarity of Black Hole Evaporation in Final-State Projection Models,''
  arXiv:1308.4209 [hep-th].

\bibitem{Hsu:2013fra} 
  S.~D.~H.~Hsu,
  ``Factorization of Unitarity and Black Hole Firewalls,''
  arXiv:1308.5686 [hep-th].

\bibitem{Hui:2013jfa} 
  L.~Hui and I-S.~Yang,
  ``Complementarity + Back-Reaction Is Enough,''
  Phys.\ Rev.\ D {\bf 89}, 084011 (2014)
  [arXiv:1308.6268 [hep-th]].

\bibitem{Mathur:2013qda} 
  S.~D.~Mathur,
  ``What Does Strong Subadditivity Tell Us about Black Holes?,''
  arXiv:1309.6583 [hep-th].

\bibitem{Giddings:2013noa} 
  S.~B.~Giddings and Y.~Shi,
  ``Effective Field Theory Models for Nonviolent Information Transfer from Black Holes,''
  arXiv:1310.5700 [hep-th].

\bibitem{Papadodimas:2013wnh} 
  K.~Papadodimas and S.~Raju,
  ``The Black Hole Interior in AdS/CFT and the Information Paradox,''
  Phys.\ Rev.\ Lett.\  {\bf 112}, 051301 (2014)
  [arXiv:1310.6334 [hep-th]].

\bibitem{Papadodimas:2013jku} 
  K.~Papadodimas and S.~Raju,
  ``State-Dependent Bulk-Boundary Maps and Black Hole Complementarity,''
  Phys.\ Rev.\ D {\bf 89}, 086010 (2014)
  [arXiv:1310.6335 [hep-th]].

\bibitem{Nomura:2013lia} 
  Y.~Nomura and S.~J.~Weinberg,
  ``The Entropy of a Vacuum: What Does the Covariant Entropy Count?,''
  arXiv:1310.7564 [hep-th].

\bibitem{Abramowicz:2013dla} 
  M.~A.~Abramowicz, W.~Klu\'{z}niak and J.-P.~Lasota,
  ``Mass of a Black Hole Firewall,''
  Phys.\ Rev.\ Lett.\  {\bf 112}, 091301 (2014)
  [arXiv:1311.0239 [gr-qc]].

\bibitem{Verlinde:2013qya} 
  E.~Verlinde and H.~Verlinde,
  ``Behind the Horizon in AdS/CFT,''
  arXiv:1311.1137 [hep-th].

\bibitem{Ilgin:2013iba} 
  I.~Ilgin and I-S.~Yang,
  ``Causal Patch Complementarity: The Inside Story for Old Black Holes,''
  Phys.\ Rev.\ D {\bf 89}, 044007 (2014)
  [arXiv:1311.1219 [hep-th]].

\bibitem{Braunstein:2013mba} 
  S.~L.~Braunstein and S.~Pirandola,
  ``Evaporating Black Holes have Leaky Horizons or Exotic Atmospheres,''
  arXiv:1311.1326 [quant-ph].

\bibitem{Susskind:2013lpa} 
  L.~Susskind,
  ``New Concepts for Old Black Holes,''
  arXiv:1311.3335 [hep-th].

\bibitem{Susskind:2014gsa} 
  L.~Susskind,
  ``The Limits of Black Hole Complementarity'' (2014).
  
\bibitem{Akhoury:2013bia} 
  R.~Akhoury,
  ``Unitary S Matrices With Long-Range Correlations and the Quantum Black Hole,''
  arXiv:1311.5613 [hep-th].

\bibitem{Susskind:2013aaa} 
  L.~Susskind,
  ``Butterflies on the Stretched Horizon,''
  arXiv:1311.7379 [hep-th].

\bibitem{Hossenfelder:2014jha} 
  S.~Hossenfelder,
  ``Disentangling the Black Hole Vacuum,''
  arXiv:1401.0288 [hep-th].

\bibitem{Brustein:2014faa} 
  R.~Brustein and A.~J.~M.~Medved,
  ``Firewalls, Smoke and Mirrors,''
  arXiv:1401.1401 [hep-th].

\bibitem{Golovnev:2014jua} 
  A.~Golovnev,
  ``Smooth Horizons and Quantum Ripples,''
  arXiv:1401.2810 [gr-qc].

\bibitem{Banks:2014xja} 
  T.~Banks, W.~Fischler, S.~Kundu and J.~F.~Pedraza,
  ``Holographic Space-time and Black Holes: Mirages As Alternate Reality,''
  arXiv:1401.3341 [hep-th].

\bibitem{Freivogel:2014dca} 
  B.~Freivogel,
  ``Energy and Information Near Black Hole Horizons,''
  arXiv:1401.5340 [hep-th].

\bibitem{Hawking:2014tga} 
  S.~W.~Hawking,
  ``Information Preservation and Weather Forecasting for Black Holes,''
  arXiv:1401.5761 [hep-th].

\bibitem{Giddings:2014nla} 
  S.~B.~Giddings,
  ``Modulated Hawking Radiation and a Nonviolent Channel for Information Release,''
  arXiv:1401.5804 [hep-th].

\bibitem{Moffat:2014eua} 
  J.~W.~Moffat,
  ``Stochastic Quantum Gravity, Gravitational Collapse and Grey Holes,''
  arXiv:1402.0906 [hep-th].

\bibitem{Lowe:2014vfa} 
  D.~A.~Lowe and L.~Thorlacius,
  ``Black Hole Complementarity: the Inside View,''
  arXiv:1402.4545 [hep-th].

\bibitem{Susskind:2014rva} 
  L.~Susskind,
  ``Computational Complexity and Black Hole Horizons,''
  arXiv:1402.5674 [hep-th].

\bibitem{Susskind:2014ira} 
  L.~Susskind,
  ``Addendum to Computational Complexity and Black Hole Horizons,''
  arXiv:1403.5695 [hep-th].
  
\bibitem{Hollowood:2014hta} 
  T.~J.~Hollowood,
  ``Schrodinger's Cat and the Firewall,''
  arXiv:1403.5947 [hep-th].

\bibitem{Sasaki:2014spa} 
  M.~Sasaki and D.-h.~Yeom,
  ``Thin-Shell Bubbles and Information Loss Problem in Anti de Sitter Background,''
  arXiv:1404.1565 [hep-th].

\bibitem{Varela:2014qua} 
  J.~Varela,
  ``Semi-Classical Field Theory as Decoherence Free Subspaces,''
  arXiv:1404.3498 [hep-th].

\bibitem{AMPSS} 
  A.~Almheiri, D.~Marolf, J.~Polchinski, D.~Stanford and J.~Sully,
  ``An Apologia for Firewalls,''
  JHEP {\bf 1309}, 018 (2013)
  [arXiv:1304.6483 [hep-th]].

\bibitem{Wheeler}
  J.~A.~Wheeler,
  {\it Relativity Groups and Topology, 1963 Les Houches Lectures},
  Gordon and Breach Science Publishers, Inc., New York, 1964.

\bibitem{Sorkin:1981wd} 
  R.~D.~Sorkin, R.~M.~Wald and Z.~J.~Zhang,
  ``Entropy of selfgravitating radiation,''
  Gen.\ Rel.\ Grav.\  {\bf 13}, 1127 (1981).

\bibitem{Hsu:2007dr} 
  S.~D.~H.~Hsu and D.~Reeb,
  ``Black Hole Entropy, Curved Space and Monsters,''
  Phys.\ Lett.\ B {\bf 658}, 244 (2008)
  [arXiv:0706.3239 [hep-th]].

\bibitem{Hsu:2008yi} 
  S.~D.~H.~Hsu and D.~Reeb,
  ``Unitarity and the Hilbert Space of Quantum Gravity,''
  Class.\ Quant.\ Grav.\  {\bf 25}, 235007 (2008)
  [arXiv:0803.4212 [hep-th]].

\bibitem{Hsu:2009kv} 
  S.~D.~H.~Hsu and D.~Reeb,
  ``Monsters, Black Holes and the Statistical Mechanics of Gravity,''
  Mod.\ Phys.\ Lett.\ A {\bf 24}, 1875 (2009)
  [arXiv:0908.1265 [gr-qc]].

\bibitem{Ong:2013mba} 
  Y.~C.~Ong and P.~Chen,
  ``The Fate of Monsters in Anti-de Sitter Spacetime,''
  JHEP {\bf 1307}, 147 (2013)
  [arXiv:1304.3803 [hep-th]].
  
\bibitem{Banks:2000fe} 
  T.~Banks,
  ``Cosmological Breaking of Supersymmetry? or Little Lambda Goes Back to the Future 2,''
  hep-th/0007146.
  
\bibitem{Bousso:2000nf} 
  R.~Bousso,
  ``Positive Vacuum Energy and the N Bound,''
  JHEP {\bf 0011}, 038 (2000)
  [hep-th/0010252].
  
\bibitem{Witten:2001kn} 
  E.~Witten,
  ``Quantum Gravity in de Sitter Space,''
  hep-th/0106109.  
  
\bibitem{Bousso:2002fi} 
  R.~Bousso, O.~DeWolfe and R.~C.~Myers,
  ``Unbounded Entropy in Space-Times with Positive Cosmological Constant,''
  Found.\ Phys.\  {\bf 33}, 297 (2003)
  [hep-th/0205080].
  
\bibitem{Page:1995dc} 
  D.~N.~Page,
  ``Sensible Quantum Mechanics: Are Only Perceptions Probabilistic?,''
  quant-ph/9506010.
  
\bibitem{Page:1995kw}
  D.~N.~Page,
  ``Sensible Quantum Mechanics: Are Probabilities Only in the Mind?,''
  Int.\ J.\ Mod.\ Phys.\ D {\bf 5}, 583 (1996)
  [gr-qc/9507024].
  
\bibitem{Page:2000cs} 
  D.~N.~Page,
  ``Can Quantum Cosmology Give Observational Consequences of Many Worlds Quantum Theory?,'' Proceedings of the 8th Canadian Conference on General Relativity and Relativistic Astrophysics, Montreal, Canada, 1999 June 10-12,
  gr-qc/0001001.  
  
\bibitem{Page:2001ba} 
  D.~N.~Page,
  ``Mindless Sensationalism: A Quantum Framework for Consciousness,''
  in {\it Consciousness: New Philosophical Perspectives}, 
  eds. Quentin Smith and Alexander Jokic 
  (Oxford, Oxford University Press, 2003), pp. 468-506,
  quant-ph/0108039.  

\bibitem{Page:2008mx} 
  D.~N.~Page,
  ``Typicality Derived,''
  Phys.\ Rev.\ D {\bf 78}, 023514 (2008)
  [arXiv:0804.3592 [hep-th]].
    
\bibitem{Page:2008ns} 
  D.~N.~Page,
  ``Insufficiency of the Quantum State for Deducing Observational Probabilities,''
  Phys.\ Lett.\ B {\bf 678}, 41 (2009)
  [arXiv:0808.0722 [hep-th]].  
  
\bibitem{Page:2009qe} 
  D.~N.~Page,
  ``The Born Rule Fails in Cosmology,''
  JCAP {\bf 0907}, 008 (2009)
  [arXiv:0903.4888 [hep-th]].  

\bibitem{Page:2011sr} 
  D.~N.~Page,
  ``Consciousness and the Quantum,''
  arXiv:1102.5339 [quant-ph]. 
  
\bibitem{Clarkson:2003kt} 
  R.~Clarkson, A.~M.~Ghezelbash and R.~B.~Mann,
  ``Entropic N Bound and Maximal Mass Conjectures Violation in Four-Dimensional Taub-Bolt(NUT)-dS Space-Times,''
  Nucl.\ Phys.\ B {\bf 674}, 329 (2003)
  [hep-th/0307059].  
  
\bibitem{Clarkson:2004yp}
  R.~Clarkson, A.~M.~Ghezelbash and R.~B.~Mann,
  ``A Review of the N-bound and the Maximal Mass Conjectures Using NUT-Charged dS Spacetimes,''
  Int.\ J.\ Mod.\ Phys.\ A {\bf 19}, 3987 (2004)
  [hep-th/0408058].

\bibitem{Giddings:2007nu} 
  S.~B.~Giddings and D.~Marolf,
  ``A Global picture of quantum de Sitter space,''
  Phys.\ Rev.\ D {\bf 76}, 064023 (2007)
  [arXiv:0705.1178 [hep-th]].
    
\bibitem{Page:2007hp} 
  D.~N.~Page,
  ``No-Bang Quantum State of the Cosmos,''
  Class.\ Quant.\ Grav.\  {\bf 25}, 154011 (2008)
  [arXiv:0707.2081 [hep-th]].

\bibitem{Page:2009ct} 
  D.~N.~Page,
  ``Symmetric-Bounce Quantum State of the Universe,''
  JCAP {\bf 0909}, 026 (2009)
  [arXiv:0907.1893 [hep-th]].

\bibitem{Ade:2013zuv} 
  P.~A.~R.~Ade {\it et al.}  [Planck Collaboration],
  ``Planck 2013 Results. XVI. Cosmological Parameters,''
  arXiv:1303.5076 [astro-ph.CO].
  
\bibitem{Scott:2013oib} 
  D.~Scott, A.~Narimani and D.~N.~Page,
  ``Cosmic Mnemonics,''
  arXiv:1309.2381 [astro-ph.CO].

\bibitem{Page:2011yd} 
  D.~N.~Page,
  ``Finite Canonical Measure for Nonsingular Cosmologies,''
  JCAP {\bf 1106}, 038 (2011)
  [arXiv:1103.3699 [hep-th]].
  
\bibitem{Hawking-2014-04-07}
  S.~W.~Hawking, private communication (2014 April 7 $\sim$ 16:30-18:00 British Summer Time).

\bibitem{Horowitz:1995ta} 
  G.~T.~Horowitz and R.~C.~Myers,
  ``The Value of Singularities,''
  Gen.\ Rel.\ Grav.\  {\bf 27}, 915 (1995)
  [gr-qc/9503062].

\bibitem{monogamy} 
  B.~C.~Sanders and J.~S.~Kim, 
  ``Monogamy and Polygamy of Entanglement in Multipartite Quantum Systems,'' 
  Appl.\ Math.\ Inf.\ Sci.\ {\bf 4}, 281-288 (2010).

\end{thebibliography}
\end{document}